\documentclass[11pt, oneside]{article}   	
\usepackage[top=25mm, bottom=25mm, right=20mm, left=20mm]{geometry} 
\usepackage{geometry}                		
\usepackage{enumitem}
\usepackage{lipsum}            		
\usepackage{graphicx}				
\usepackage[%
    colorlinks=true,
    pdfborder={0 0 0},
    linkcolor=red!70!violet
]{hyperref}				
\usepackage{amssymb}
\usepackage{amsmath}
\usepackage{graphicx,subfigure}
\usepackage{tikz}

\usepackage{physics}
\usepackage{blindtext}
\usepackage[T1]{fontenc}
\usepackage[utf8]{inputenc}			
\usepackage[toc,page]{appendix}
\usepackage{amssymb}
\usepackage{hyperref}
\hypersetup{
    colorlinks=true,
    linkcolor=red!70!violet,
    filecolor=magenta,      
    urlcolor=red!40!violet,
}
\usepackage{hyperref}
\usepackage[utf8]{inputenc}
\usepackage{algorithm}
\usepackage{algpseudocode}
\usepackage{CormorantGaramond}
\let\oldnormalfont\normalfont
\def\normalfont{\oldnormalfont\mdseries}
\usepackage{xcolor}
\usepackage{sectsty}
\usepackage{caption}
\usepackage{multirow}
\usepackage{multicol}

\title{High Luminosity LHC data collected by CMS experiment - an excellent ground for the search of Rare Radiative $\mathrm{B_s^0}$ meson decays : A Review }
\author{Alibordi Muhammad\\Faculty of Physics\\ University of Warsaw\\ Ludwika Pasteura 5, 02-093 , Warsaw, Poland}
\date{September 2025}

\begin{document}

\maketitle
\begin{abstract}
The High-Luminosity Large Hadron Collider (HL-LHC) era presents unprecedented opportunities for discovering rare radiative decays in the flavor sector, particularly through the Compact Muon Solenoid (CMS) experiment. This review examines the feasibility and methodology for searching rare $\mathrm{B_s^0}$ meson decays involving photons, with emphasis on the theoretically and experimentally challenging channel $\mathrm{B_s^0 \to \mu^+\mu^-\gamma}$. Such flavor-changing neutral current processes, forbidden at tree level in the Standard Model and proceeding via loop-induced penguin diagrams, provide unique sensitivity to Wilson coefficients $C_{7\gamma}$, $C_{9V}$, and $C_{10A}$—key parameters in the effective Hamiltonian formalism that bridge perturbative and non-perturbative QCD dynamics. The primary experimental challenge lies in low-$p_T$ photon reconstruction ($2$--$20$ GeV) amid extreme pile-up conditions ($\langle \mu \rangle \sim 60$ at HL-LHC). We demonstrate that collider data intrinsically inhabit curved statistical manifolds shaped by conservation laws, detector geometry, and kinematic constraints—structure invisible to conventional Euclidean algorithms. The Fisher-Rao information metric quantifies this geometry through sectional curvature: negative (hyperbolic) in regions dominated by hierarchical QCD backgrounds, positive (spherical) near kinematic boundaries. Respecting this geometric structure through curvature-aware analysis techniques offers a pathway beyond traditional approaches. This review develops a possible framework connecting hierarchy of collider data, information geometry together with quantum machine learning. These geometric and quantum perspectives are exploratory, not established, but highlight methodological directions worth considering in the HL-LHC era.
\end{abstract}
\newpage
\tableofcontents
\newpage
\section{Prologue}
The $\mathrm{B_s^0}$-meson system is a di-quark composite, often regarded as a gold standard for studies of CP-violation~\cite{rmpbs1}. A detailed account of the CP-violating(see Ap.~\ref{ap:cp}) properties of the $\mathrm{B_s^0}$ system can be found in Refs.~\cite{ambs3,ambs4,ambs5}. The measured values of key $\mathrm{B_s^0}$ meson properties, as reported by the Particle Data Group~\cite{pdgbs2}, are summarized in Tab.~\ref{tab:Bs}. For the most recent global fits, readers are encouraged to consult the Heavy Flavor Averaging Group (HFLAV) website\footnote{See \href{https://hflav.web.cern.ch}{HFLAV website}}, and for the inputs to the European Strategy for Particle Physics, see Ref.~\cite{flavbs6}. Building on this foundation, the present handout turns toward a more forward-looking question: the possibilities of rare radiative decays of the $\mathrm{B_s^0}$ meson. These channels, though experimentally challenging, offer unique sensitivity to physics beyond the Standard Model (SM) and serve as a testing ground for new ideas in flavor dynamics.
\begin{table}[htbp]
\centering
\begin{tabular}{l l}
\hline
 Properties  & Corresponding Values  \\ 
 \hline
$\mathrm{m_{B_s^0}}$ & $5366.91\pm 0.11\, \text{MeV}$ \\
$\tau_\mathrm{B_s^0}$ & $(1.516\pm 0.006)\times 10^{-12}\, s$ \\
$\Delta\Gamma_\mathrm{B_s^0}$ & $(80 \pm 5)\times 10^9 \, s^{-1}$ \\
$\Delta m_\mathrm{B_s^0}$ & $(17.765\pm 0.005)\times 10^{12}\,\mathrm{\hbar s}^{-1}$ \\
$\beta_s(b\to c\bar{c}s)$ & $0.019\pm 0.008\,\text{rad}$\\
$\abs{\lambda}(b\to c\bar{c}s)$ & $0.989\pm 0.008$ \\
 \hline
\end{tabular}
\caption{Properties of $\mathrm{B_s^0}$ Meson are taken from particle data group~\cite{pdgbs2}.}
\label{tab:Bs}
\end{table}
Rare radiative decays via Flavor Changing Neutral Current (FCNC) through the quark transition ($b\to s\gamma, b\to d\gamma$) are forbidden at the tree level of the SM, though accessible via the penguin or box diagram~\cite{pdgbs2}. Because of their low branching fractions, these decay modes demand very large event samples, which in practice means high-luminosity Large Hadron Collider (LHC) data. During the ongoing Run3 data taking period, all LHC based experiments (see Sec.~\ref{sec:cms}) are intend to collect high luminosity data (as for CMS the target is 300fb$^{-1}$), which provides an excellent opportunity to utilize such data by implementing state-of-the-art analysis technique in order to find the signature of rare radiative decays.The central question motivating this handout is therefore: \textit{Why is now the best time to pursue rare radiative decays?}. After a brief discussion of the theoretical background in the Sec.~\ref{sec:th}, I will try to understand the current situation of the CMS experiment in the Sec.~\ref{sec:cms}. As this is my 10$^{\text{th}}$ year in the CMS collaboration, the perspective presented here is naturally CMS-centric. Then I will give an overview of the feasibility checks connecting theory to experiment in the Sec.~\ref{sec:fc}. After that I will move to most important part of the discussion, which is the intrinsic hierarchy, and the  curvature awareness of the collider data in Sec.~\ref{sec:hcd}. Finally we will summaries the discussion via looking at the possibilities of applying a curvature aware quantum machine learning model for analyzing high luminosity data in the search of rare radiative decays in Sec.~\ref{sec:bcqml}.

\section{Theoretical Background}
\label{sec:th}
\subsection{Effective Hamiltonian}
\begin{figure}
    \centering
    \includegraphics[width=0.5\linewidth]{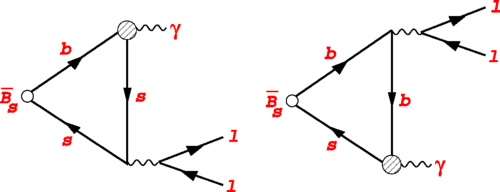}
    \caption{Penguin diagrams responsible for the virtual photon emission and the Wilson coeffcient $C_{7\gamma}$ appear in the Eq. ~\ref{ref:eh} for the $\mathrm{B_s^0\to s\ell^+\ell^-}$ decay~\cite{rad3}.}
    \label{fig:pd}
\end{figure}

I want to consider cases with direct three body decays ($\mathrm{B_s^0\to\mu^+\mu^-\gamma, B_d^0\to\mu^+\mu^-\gamma,}$) and decays through resonances ($\mathrm{B_s^0\to J/\psi(\mu^+\mu^-)\gamma, B_s^0\to K^{*0}(\mu^+\mu^-)\gamma}$). The FCNC Effective Hamiltonian~\cite{rad3} (Eq.~\ref{ref:eh}) for the $b\to s\ell^+\ell^-$ decays provided by the Wilson Operator Product Expansion(OPE) at the scale $\Lambda$ is given by, 
\begin{equation}
\begin{split}
\mathcal{H}_\text{eff}^{b\to s\ell\ell}  &=\frac{G_F}{\sqrt{2}}\frac{\alpha_{em}}{2\pi}V_{tb}V_{ts} [-2im_b\frac{{\color{blue}C_{7\gamma}}( \Lambda)}{q^2}\cdot \bar{s}\sigma_{\mu\nu}q^\nu(1+\gamma^5)b\cdot \bar{\ell}\gamma^\mu\ell   \\
&+{\color{red}C_{9V}^\text{eff}}( \Lambda, q^2).\bar{s}\gamma^\mu(1-\gamma_5)b\cdot \bar{\ell}\gamma^\mu\ell +{\color{red}C_{10A}^\text{eff}}( \Lambda).\bar{s}\gamma^\mu(1-\gamma_5)b\cdot \bar{\ell}\gamma^\mu\gamma_5\ell],
\end{split}
\label{ref:eh}
\end{equation}
where, $C_i$'s are Wilson coefficients, $s,b,t$ denotes quark spinors, and $\ell\in(e, \mu)$.  Radiative decays are sensitive to the Wilson coefficient $C_{7\gamma}$ in addition to the coefficients $C_{10A}, C_{9V}$. $C_{7\gamma}$, emerging from the virtual photon emitted in the penguin diagrams depicted in the Fig.~\ref{fig:pd}.

\subsection{Wilson Coefficients and Hadronic Dynamics}
Rare radiative and semileptonic decays of neutral $B$ mesons provide one of the sharpest laboratories for probing the interplay between short-distance physics, encoded in Wilson coefficients of the effective Hamiltonian, and long-distance hadronic dynamics that must be captured through form factors, resonant amplitudes, and non-perturbative inputs. In particular, the channels $\mathrm{B_s^0\to\mu^+\mu^-\gamma}$ and $\mathrm{B_d^0\to\mu^+\mu^-\gamma}$ represent direct three-body final states where the virtual photon couples directly to the quark-level transition $b\to s \ell^+\ell^-$~\cite{rad1,rad2,rad3}, while processes such as $\mathrm{B_s^0\to J/\psi(\mu^+\mu^-)\gamma}$ or $\mathrm{B_s^0\to K^{*0}(\mu^+\mu^-)\gamma}$ involve the formation of intermediate hadronic resonances that subsequently decay to a dilepton system~\cite{rad4,rad5,rad6}. Both categories are sensitive to the fundamental coefficients $C_{7\gamma}$, $C_{9V}$, and $C_{10A}$ defined in Eq.~\ref{ref:eh}, but the manner in which these coefficients manifest in observable distributions differs substantially between non-resonant and resonant channels.  In the direct three-body decays, the amplitude is proportional to the short-distance coefficient $C_{7\gamma}$ that governs the electromagnetic dipole operator, together with $C_{9V}$ and $C_{10A}$ which encode the vector and axial-vector couplings of the dilepton current~\cite{rad9,rad10,rad11,rad12,rad13}. The relevant hadronic physics enters through transition form factors, typically parameterized as $\mathrm{\langle \gamma | \bar{s} \Gamma b | B_s^0\rangle}$, where $\Gamma$ denotes the Lorentz structure of the effective operator. These form factors must be obtained from non-perturbative methods such as QCD sum rules or lattice QCD~\cite{dg2,dg4}, and the associated uncertainties limit the achievable precision, yet the relative absence of resonant contamination makes these channels comparatively clean probes of short-distance coefficients. In contrast, resonant modes such as $\mathrm{B_s^0\to J/\psi(\mu^+\mu^-)\gamma}$ or $\mathrm{B_s^0\to K^{*0}(\mu^+\mu^-)\gamma}$ proceed through intermediate on-shell states, most notably charmonium resonances in the case of $\mathrm{J/\psi}$ and higher vector mesons in the case of $\mathrm{K^{*0}}$. In these decays, the effective Wilson coefficient $C_{9V}$ is modified by long-distance contributions that can be parameterized as a $q^2$-dependent shift,
\begin{equation}
C_{9V}^{\mathrm{eff}}(q^2) = C_{9V} + Y(q^2),
\end{equation}
where $Y(q^2)$ accounts for charm-loop effects and direct resonance contributions~\cite{rad4,rad7,rad8,rad13}. This functional dependence on the dilepton invariant mass squared leads to pronounced peaks in the differential branching ratio whenever the invariant mass approaches the mass of a narrow resonance such as $\mathrm{J/\psi}$ or $\psi(2S)$. Experimentally, these peaks dominate the dilepton spectrum and must be either excluded or carefully modeled when extracting information about the underlying short-distance coefficients. The distinction between non-resonant and resonant modes is thus essential: while the former offer direct though rare access to short-distance coefficients with limited contamination, the latter provide large event samples but suffer from substantial theoretical uncertainties due to hadronic dynamics. From the perspective of effective field theory, both cases remain governed by the same Hamiltonian structure, but the non-local hadronic matrix elements differ sharply, requiring distinct treatment in phenomenological analyses. A unifying feature is that both rely critically on our ability to calculate or parameterize hadronic form factors, a problem intrinsically tied to the non-perturbative regime of QCD~\cite{rad4,rad8}. At a deeper level, the very existence of a discrete spectrum of hadrons and narrow resonances reflects confinement, the property that quarks and gluons cannot be isolated at low energies. This feature is formally related to the mass gap problem in Yang–Mills theory, one of the Clay Institute Millennium Prize problems, which seeks to prove that non-Abelian gauge theories such as QCD exhibit a finite lowest excitation energy above the vacuum. While the study of $\mathrm{B}$ decays does not solve this mathematical conjecture, it is nevertheless true that all of the hadronic structures entering the amplitudes—from the $\mathrm{B}$ meson itself to the $\mathrm{J/\psi}$ resonance—are physical manifestations of confinement and of the non-zero mass gap.  In practice, this means that phenomenologists must work with effective coefficients that mix perturbative short-distance physics, well-described by the operator product expansion, with phenomenological models of non-perturbative dynamics that stem from confinement~\cite{rad10,rad11,rad12}. This tension between clean short-distance inputs and messy long-distance effects is precisely what makes rare $\mathrm{B}$ decays both challenging and powerful probes: small deviations from SM expectations may indicate new heavy particles, but only if the hadronic physics is sufficiently under control. Recent global fits to $b\to s \ell^+\ell^-$ processes have suggested anomalies in observables such as $R_K$ and $R_{K^*}$, which are ratios of branching fractions to muon versus electron pairs~\cite{flavbs6}. While these anomalies are not directly in the radiative channels under discussion, they illustrate the same sensitivity of $C_{9V}$ and $C_{10A}$ to potential lepton-flavor universality violation. Radiative decays provide a complementary window, especially since they are strongly sensitive to $C_{7\gamma}$, and thus can help disentangle patterns of new physics couplings. For instance, a deviation in the photon-energy spectrum of $\mathrm{B_s^0\to\mu^+\mu^-\gamma}$ relative to SM predictions would directly constrain dipole-type operators, while shifts in angular observables of resonant modes would more cleanly test modifications of $C_{9V}$ and $C_{10A}$\cite{dg3,dg5,dg6}. From a methodological standpoint, this underscores the importance of combining lattice QCD determinations of form factors, dispersive bounds, and experimental data-driven approaches in order to reduce theoretical uncertainties~\cite{rad3,rad4,rad8}. Furthermore, the treatment of resonant contributions requires sophisticated modeling of interference effects, as the narrow-width approximation is not always sufficient. Beyond purely phenomenological considerations, there is also a conceptual payoff in studying these systems: they embody the marriage of local quantum field theory, in which the effective Hamiltonian is rigorously defined, with emergent hadronic degrees of freedom that cannot be derived perturbatively but must instead be inferred from non-perturbative dynamics. In this sense, the analysis of radiative $\mathrm{B}$ decays is not only about testing for new physics but also about deepening our empirical handle on how confinement and the hadronic spectrum imprint themselves on weak decays. While the mass gap conjecture~\cite{mg} remains unsolved, the practical reality that we observe well-defined hadronic excitations with finite gaps above the vacuum is the very reason why the effective field theory approach is viable. The Wilson coefficients are universal and perturbatively calculable, but their translation into observables depends on the structured, gapped spectrum of QCD bound states. Thus, every measurement of a rare $\mathrm{B}$ decay simultaneously tests the SM at short distances and reminds us of the unsolved problem of confinement at long distances. The study of both non-resonant and resonant radiative $\mathrm{B}$ decays offers a uniquely rich framework for probing short-distance electroweak dynamics, testing the universality of Wilson coefficients across decay channels, and confronting the unavoidable non-perturbative effects of QCD. By treating direct three-body modes and resonant channels on the same theoretical footing—while carefully accounting for their differing hadronic systematics—one can construct a coherent global picture of the flavor sector, sensitive both to potential new physics and to the underlying structure of the strong interaction.

\subsection{Existing numeric}
Due to the presence of the photon in the final state radiative decays under question are not helicity suppressed, making it less rare than leptonic decay, such as, $\mathrm{B_s^0\to\mu^+\mu^-}$. Simultaneously, the low momentum photon in the initial state radiation(ISR) or in final state radiation(FSR) making the detectability more challenging, specially in a high energy scenario( pile-up ($\sim 60$)). I prepared a table (Tab.~\ref{tab:rd}) 
\begin{table}[htbp]
\centering
\begin{tabular}{l l}
\hline
 Decay Channel  & Branching Fraction or Upper Limit (90\% CL)  \\ 
 \hline
 $\mathrm{B_s^0 \to \mu^+\mu^-\gamma}$(LHCb) & $< 4.2 \times 10^{-8},\; m(\mu^+\mu^-) \in [2m_\mu, 1.70] \text{ GeV}/c^2$ \\
& $< 7.7 \times 10^{-8},\; m(\mu^+\mu^-) \in [2m_\mu, 2.88] \text{ GeV}/c^2$ \\
 & $< 4.2\times 10^{-8},m(\mu^+\mu^-)\in [3.92, m_{B_s^0}] \text{ GeV}/c^2$\\  
 $\mathrm{B_d^0 \to \mu^+\mu^-\gamma}$(th) & $1.31\times 10^{-10}, \text{at}\,E^\gamma_\text{min}=80\,\text{MeV}$~\cite{rad2} \\ 
$\mathrm{B_d^0 \to \ell^+\ell^-\gamma}$(th) & $(1.02\pm 0.15\pm 0.05)\times 10^{-11}, \text{at}\,q^2\in[1,6]\,\text{GeV}^2$~\cite{rad3} \\
$\mathrm{B_s^0 \to \ell^+\ell^-\gamma}$(th) & $(6.01\pm 0.08\pm 0.70)\times 10^{-9}, \text{at}\,q^2\in[1,6]\,\text{GeV}^2$~\cite{rad3} \\
 \hline 
 $\mathrm{B_s^0 \to J/\psi(\mu^+\mu^-)\gamma}$(exp) & $<7.3\times 10^{-6}$~\cite{pdgbs2}  \\
  $\mathrm{B_s^0 \to \phi \gamma}$ & $(39.4\pm 10.7\pm 5.3)\times 10^{-6}$~\cite{rad8} \\ 
 $\mathrm{B_s^0 \to \phi(\mu^+\mu^-)\gamma}$ & -- (doubly suppressed)~\cite{rad7} \\ 
 $\mathrm{B_d^0\to\rho(770)^0(\pi^+\pi^-)\gamma}$ & $(7.9\pm 0.3\pm 0.2\pm 0.2)\times 10^{-6}$ ~\cite{LHCb3} \\ 
 $\mathrm{B_d^0 \to K^{*0}\gamma}$ & $(4.39\pm 0.41\pm 0.27)\times10^{-5}$ ~\cite{rad5,rad6}\\
 $\mathrm{B_s^0 \to K^{*0}(\mu^+\mu^-)\gamma}$ & -- \\ 
 $\mathrm{B_d^0 \to K^{*0}(\mu^+\mu^-)\gamma}$ & -- \\ 
 $\mathrm{B_s^0 \to K^{*0}(K^+\pi^-)\gamma}$ & -- \\ 
 $\mathrm{B_s^0 \to \phi_{\text{inc}}(a_{\text{trk/h}}^+ a_{\text{trk/h}}^-)\gamma}$ & -- \\
 $\mathrm{B_s^0 \to \phi(K^+K^-)\gamma}$(exp) & $(3.4\pm 0.4)\times 10^{-5}$~\cite{pdgbs2}\\
 \hline
\end{tabular}
\caption{List of rare radiative and related decay channels and their current experimental limits or branching fractions.}
\label{tab:rd}
\end{table}
demonstrating available values of the branching fraction corresponding to various rare radiative decays. The empty spaces shows the value is not available neither experimentally nor theoretically. As for the decay channel $\mathrm{B_s^0 \to \phi(\mu^+\mu^-)\gamma}$, it is strongly suppressed due to the small branching fraction of the subprocess $\phi(1020)\to \mu^+\mu^-$, which is of the order $\mathcal{O}(10^{-4})$. Consequently, the overall process becomes doubly rare, as it is proportional to the product of the radiative penguin branching fraction and the tiny $\mathcal{B}(\phi(1020)\to \mu^+\mu^-)$~\cite{pdgbs2}. Franz and collaborators have investigated various properties of the decay $\mathrm{B_s^0 \to \phi\gamma}$, but without specifying details regarding the dimuon final state from $\phi(1020)\to \mu^+\mu^-$~\cite{rad7}. The LHCb collaboration has published two analyses of the radiative mode $\mathrm{B_s^0\to\mu^+\mu^-\gamma}$, reporting upper limits on the branching fraction~\cite{LHCb1, LHCb2}. The latest results, obtained in three dimuon mass bins, are already summarized in the first row of Tab.~\ref{tab:rd}. These studies employed photons with transverse momentum $p_T$ just above $1\,\text{GeV}$. In their strategy, $\mathrm{B_s^0\to\phi(K^+K^-)\gamma}$ was used as the control channel, while $\mathrm{B_s^0\to J/\psi(\mu^+\mu^-)\eta}$ served as the normalization channel. Additionally, LHCb has studied the radiative decay $\mathrm{B_d^0\to\rho(770)^0(\pi^+\pi^-)\gamma}$, with the results also listed in Tab.~\ref{tab:rd}. Even though not directly related to the dimuon radiative modes, the amplitude analyses of $\mathrm{B_s^0 \to K^+K^-\gamma}$ and $\mathrm{\Lambda_b^0 \to pK^-\gamma}$~\cite{LHCb4,LHCb5}, together with the photon polarization study in $\mathrm{B_s^0 \to \phi e^+e^-}$~\cite{LHCb4}, have provided valuable insights into the accessible kinematic phase space and the dynamics of radiative penguin transitions.

\section{Compact Muon Solenoid}
\subsection{Status}
\label{sec:cms}
The Compact Muon Solenoid (CMS) experiment at the Large Hadron Collider (LHC) was originally conceived with the broad ambition of being a general-purpose detector~\cite{cms3}, yet its particular strengths have rendered it uniquely powerful in flavor physics and rare decay searches. At first glance, CMS is often associated with high-$p_T$ phenomena such as Higgs boson discovery or searches for supersymmetry. However, the same design elements that make CMS robust in the multi-TeV regime also confer advantages in the sub-10 GeV domain relevant to heavy-flavor decays. In particular, the high-granularity electromagnetic calorimeter (ECAL)~\cite{cms3}, the precision silicon tracker~\cite{cms4}, and the nearly hermetic muon system~\cite{cms1} collectively make CMS one of the few experiments that can reconstruct final states combining very soft photons with high-quality muons. This balance between reach and precision underpins the experiment's capacity to address questions such as rare radiative decays of the $\mathrm{B_s^0}$ meson. The central feature of CMS is its 3.8 T superconducting solenoid~\cite{cms3}, which allows for fine tracking resolution even for low-$p_T$ particles. This is essential in the context of rare radiative decays: photons are often reconstructed through their conversions into electron–positron pairs, where the precision of the silicon tracker~\cite{cms4} directly controls the momentum resolution of the reconstructed photon candidate. Similarly, the performance of the muon chambers~\cite{cms1} ensures clean triggering on dimuon final states, a prerequisite for capturing $\mathrm{B_s^0 \to \mu^+\mu^-\gamma}$~\cite{ambs5} in real time. Indeed, much of the knowledge accumulated during the $\mathrm{B_s^0 \to \mu^+\mu^-}$ analyses has prepared the ground for these more elusive channels, showing that the synergy of tracking, calorimetry, and muon systems can sustain analyses with signal yields numbering in the tens of events. Another remarkable feature of CMS is its data acquisition and trigger system~\cite{cms2}, which has undergone continuous upgrades to adapt to the extreme pile-up environment of the High-Luminosity LHC (HL-LHC). Rare decay searches sit at the edge of trigger feasibility: the dimuon triggers used for flavor physics must maintain efficiency down to a few GeV in muon $p_T$, while not overwhelming the bandwidth with background from heavy-flavor production. Recent developments in machine-learning–based trigger algorithms suggest that the Level-1 (L1) and High-Level Trigger (HLT) systems may evolve into more adaptive filters, capable of recognizing rare decay topologies at the earliest stages of event selection~\cite{cms6}. For $\mathrm{B_s^0 \to \mu^+\mu^-\gamma}$~\cite{ambs5}, the ability to incorporate information about displaced vertices, low-$p_T$ tracks, and calorimeter clusters into L1 decisions could be the deciding factor for collecting a sufficient number of events~\cite{eps}. Historically, CMS has demonstrated that it can pivot into rare decay physics with agility. The observation of $\mathrm{B_s^0 \to \mu^+\mu^-}$~\cite{ambs5}, achieved jointly with LHCb, remains a hallmark of this versatility. In that case, CMS contributed crucially despite not being a dedicated flavor experiment, leveraging the immense luminosity and excellent muon identification capabilities~\cite{cms1} to stand on equal footing with LHCb. The present situation is similar: while LHCb remains the natural leader for rare decay searches due to its forward geometry and optimized tracking, CMS can provide complementary results with different systematics, validating or challenging potential anomalies. This complementarity becomes particularly important for radiative channels, where the combination of different detector acceptances and reconstruction techniques helps to disentangle detector-specific effects from genuine physical signals. The CMS Phase-2 upgrade program~\cite{cms8} ensures that the detector remains competitive in rare decay searches through the HL-LHC era~\cite{cms5}. The upgraded tracker~\cite{cms7} will provide extended coverage and improved material mapping, directly benefiting photon conversion reconstruction. The high-granularity calorimeter~\cite{cms8} in the endcaps (HGCAL) will sharpen the ability to distinguish electromagnetic showers from pile-up, a feature that may turn out to be decisive in identifying soft photons in dense environments. Meanwhile, advances in muon detector coverage and electronics will allow CMS to sustain low-$p_T$ dimuon triggers with high efficiency~\cite{cms6}. Together, these developments establish CMS as a long-term player in the study of rare radiative decays, not merely as an auxiliary contributor but as a laboratory where advanced data analysis methods can be stress-tested under extreme luminosity conditions. The CMS experiment occupies a unique position at the intersection of high-statistics collider physics and high-precision flavor measurements. Its technical design~\cite{cms3}, its experience with rare processes, and its upgrade trajectory~\cite{cms5,cms6,cms7} converge to make it a fertile ground for exploring $\mathrm{B_s^0 \to \mu^+\mu^-\gamma}$ and related decays. While it may not match the specialized coverage of LHCb, CMS brings to the table luminosity, complementary acceptance, and the capacity to innovate in trigger and reconstruction~\cite{cms8,cms2,cms4}. These strengths position CMS not only to confirm potential signals but also to pioneer the methodological frontiers—particularly in the integration of machine learning, information geometry, and quantum-inspired analysis techniques—that may define the next generation of rare decay studies.

\subsection{Feasibility Check For Radiative Decays}
\label{sec:fc}
The feasibility of studying rare radiative $\mathrm{B}$ decays at CMS hinges on the careful interplay between detector performance, statistical power, and computational innovation. From the production side, the Large Hadron Collider (LHC) provides an extraordinarily favorable environment: with its large $\mathrm{b\bar{b}}$ cross section, billions of $\mathrm{B}$  mesons are produced within the CMS acceptance during each run. At the HL-LHC, with a projected dataset exceeding $3\,\mathrm{ab}^{-1}$, even decays with branching fractions of $\mathcal{O}(10^{-9})$ yield event counts that are, at least statistically, accessible. The real challenge lies not in the yield but in the reconstruction of diverse final states in an environment dominated by background. For radiative channels, whether they involve purely leptonic ($\mathrm{B_{s,d}^0 \to \ell^+\ell^-\gamma}$), semileptonic ($\mathrm{B_{s,d}^0 \to V(\ell^+\ell^-)\gamma}$), or hadronic ($\mathrm{B_{s,d}^0 \to V(h^+h^-)\gamma}$) final states, the photon typically emerges with transverse momentum in the range $2$–$20\,\text{GeV}$—precisely the region where standard photon identification techniques face their greatest challenges. Photon reconstruction at low $p_T$ remains one of the most difficult tasks in hadron collider physics. In CMS, the electron and photon reconstruction framework~\cite{ph1} relies on energy deposits in the electromagnetic calorimeter (ECAL) crystals, where deposits above an electronic threshold are clustered to form super-clusters. These super-clusters are designed to recover energy losses from bremsstrahlung and photon conversions, with photons identified as clusters not matched to charged-particle tracks in the silicon tracker. While this framework has demonstrated excellent performance for high-$p_T$ photons~\cite{ph2}, particularly those from Higgs or electroweak processes, it becomes less effective in the low-$p_T$ regime critical for $\mathrm{B}$ physics. Current isolation-based identification criteria are optimized for $p_T^\gamma > 10\,\text{GeV}$, leaving a gap for the soft-photon region relevant to radiative $\mathrm{B}$ decays. Recent developments in deep learning techniques for ECAL energy clustering~\cite{ph3} and end-to-end photon reconstruction~\cite{ph4} offer promising avenues for extending photon identification to lower momenta. The optimization of high-energy photon identification~\cite{ph5} provides a foundation that can be adapted for the specific challenges of rare decay searches. The reconstruction capabilities for various final state particles present different levels of complexity. Muonic final states benefit from CMS's well-established muon system performance, as demonstrated in measurements like $\mathrm{B_s^0 \to \mu^+\mu^-}$~\cite{dg1}. For hadronic final states such as $\mathrm{B_s^0 \to \phi(K^+K^-)\gamma}$ or the unexplored $\mathrm{B_s^0 \to K^{*0}(K^+\pi^-)\gamma}$, the challenge shifts to efficient track reconstruction and particle identification in the tracker. The combination of these capabilities allows CMS to pursue a comprehensive program of radiative decay measurements, spanning the full spectrum from purely leptonic to fully hadronic channels listed in Table~\ref{tab:rd}. The statistical reach can be assessed through both direct production calculations and normalization-based approaches. For a direct estimate, taking $\mathrm{B_s^0 \to \mu^+\mu^-\gamma}$ as a benchmark case, the number of reconstructed events is given by:
\begin{equation}
\mathrm{N = \mathcal{L} \times \sigma(pp\to B_s^0+X) \times \mathcal{BR}(B_s^0\to\mu^+\mu^-\gamma) \times \mathcal{A} \times \epsilon }
\end{equation}
where $\mathcal{L} \sim 300\,\mathrm{fb}^{-1}$ represents the integrated luminosity for Run 3. The effective $B_s^0$ production cross section can be estimated as:
\begin{equation}
\begin{split}
\mathrm{\sigma(pp\to B_s^0+X)} &= \mathrm{\sigma(pp\to b\bar{b})\times 2\times f_{\mathrm{b\to B_s^0}}}\\
&\mathrm{\approx 0.5\times 10^{12}\,\text{fb} \times 2\times 0.1}\\
& \mathrm{\approx 10^{11}}\,\text{fb}
\end{split}
\end{equation}
where the factor of 2 accounts for both $\mathrm{B_s^0}$ and $\mathrm{\bar{B}_s^0}$ production. Using conservative estimates for the geometric and kinematic acceptance ($\mathcal{A} \sim 0.002$) and the combined reconstruction and selection efficiency ($\epsilon \sim 0.2$), along with theoretical branching fraction predictions of $\mathcal{BR} \sim 6 \times 10^{-9}$ from Table~\ref{tab:rd}, yields approximately 72 signal events for Run 3. This suggests that even with modest performance, observation becomes feasible with the full HL-LHC dataset of $3\,\mathrm{ab}^{-1}$, which would scale the yield to several hundred events. A complementary approach leverages normalization channels to reduce systematic uncertainties. For rare radiative decays, the signal yield can be extracted through:
\begin{equation}
\mathrm{N_\text{sig}= N_\text{norm}\times \frac{f_q}{f_q} \times \frac{1}{\mathcal{BR}(B^0_s\to J/\psi\eta)} \times \frac{\mathcal{A}_{B_s\to\mu\mu\gamma}\times \epsilon_{2\mu}\times \epsilon_{\gamma}}{\mathcal{A}_{B_s\to J/\psi\eta}\times \epsilon_{J/\psi\eta}}\times \mathcal{BR}(B_s^0\to\mu^+\mu^-\gamma)}
\end{equation}
where $N_\text{norm}$ is obtained from well-established channels like $\mathrm{B^0_s\to J/\psi\eta}$. This normalization method cancels many systematic uncertainties related to luminosity, $b$-quark production, and fragmentation. The relative acceptances and efficiencies are determined from Monte Carlo(MC) simulations, with $\mathcal{A}_{\mathrm{B_s\to\mu\mu\gamma}}\times \epsilon_{2\mu}\times \epsilon_{\gamma}$ extracted from signal MC or control channels like $\mathrm{B_s^0\to \phi(\mu^+\mu^-)\gamma}$, while $\mathcal{A}_{\mathrm{B_s\to J/\psi\eta}}\times \epsilon_{\mathrm{J/\psi\eta}}$ comes from the normalization channel MC. Trigger efficiencies require careful calibration between data and simulation. The data-to-MC correction factors are obtained through:
\begin{equation}
\mathrm{\epsilon_\text{TrigData}^{B_s\to\mu\mu\gamma}=\frac{\epsilon_\text{TrigData}^{B_s\to J/\psi\eta}}{\epsilon_\text{TrigMC}^{B_s\to J/\psi\eta}}\times \epsilon_\text{TrigMC}^{B_s\to\mu\mu\gamma}}
\end{equation}
Additional scale factors for low-momentum photon reconstruction and muon identification must be derived from control samples to ensure accurate efficiency estimates. Control channels play a crucial role in establishing the feasibility of rare radiative measurements. High-statistics modes like $\mathrm{B_s^0 \to \phi(K^+K^-)\gamma}$ with $\mathcal{BR} \sim 3.4 \times 10^{-5}$ and $\mathrm{B_d^0 \to K^{*0}\gamma}$ with $\mathcal{BR} \sim 4.4 \times 10^{-5}$ provide ideal benchmarks for validating photon reconstruction algorithms and understanding systematic uncertainties. These channels, occurring at rates several orders of magnitude higher than the ultra-rare modes, yield $\mathcal{O}(10^5)$ events even with comparable efficiencies, allowing for detailed studies of photon efficiency, energy resolution, and background rejection in the same kinematic regime. Similarly, $\mathrm{B_d^0 \to \rho^0(\pi^+\pi^-)\gamma}$ serves as an additional calibration channel for understanding hadronic resonance reconstruction in the presence of radiative photons. The statistical reach of the HL-LHC program enables a tiered approach to radiative decay measurements. Channels with branching fractions at the $10^{-5}$ level can be measured with high precision, providing stringent tests of QCD calculations and form factors. Moving to the $10^{-7}$–$10^{-8}$ regime, modes like $\mathrm{B_{s,d}^0 \to \mu^+\mu^-\gamma}$ become statistically accessible with the full dataset, though they require significant advances in background suppression. For the most suppressed channels with theoretical predictions at the $10^{-10}$–$10^{-11}$ level, even the HL-LHC may only provide upper limits, but these limits themselves constrain new physics scenarios. The heterogeneity of final states in radiative decays demands flexible analysis strategies. Machine learning approaches, already integral to CMS workflows~\cite{ph3,ph4}, offer natural solutions for handling the diverse topologies. Recent advances in geometric machine learning and information-theoretic approaches provide new tools for extracting signals from high-dimensional backgrounds. These methods are particularly valuable when dealing with mixed final states—for instance, distinguishing $\mathrm{B_s^0 \to K^{*0}(\mu^+\mu^-)\gamma}$ from $\mathrm{B_s^0 \to K^{*0}(K^+\pi^-)\gamma}$ requires simultaneous optimization of lepton identification, hadron discrimination, and photon reconstruction. Taken together, these considerations point to a realistic path forward for a comprehensive radiative decay program at CMS. The combination of HL-LHC luminosity, upgraded detector capabilities~\cite{ph1,ph2}, and advanced analysis techniques places multiple radiative channels within experimental reach. The feasibility is not uniform across all modes—hadronic channels with higher branching fractions offer immediate opportunities, while ultra-rare leptonic modes represent longer-term goals requiring the full HL-LHC dataset and methodological innovations. Success in this domain would establish CMS as a key player in the global effort to map the landscape of radiative $B$ decays, complementing dedicated flavor experiments with unique acceptance and systematic uncertainties. The program should be viewed not as pursuit of individual channels in isolation, but as a coordinated exploration of the full spectrum of radiative phenomena accessible at a high-luminosity hadron collider.

\section{Hierarchy of Collider Data : Intrinsic Curvature Awareness}
\label{sec:hcd}

The feasibility check outlined in Section~\ref{sec:fc} demonstrates that rare processes such as $\mathrm{B_s^0 \to \mu^+\mu^-\gamma}$ lie at the very edge of what is experimentally possible. Yet feasibility alone is not sufficient; the success of such searches depends on the deeper structure of the data itself. Collider data are not distributed on a flat Euclidean canvas. Rather, they occupy a hierarchy of manifolds, shaped by conservation laws, detector geometry, and probabilistic constraints. This section explores how awareness of intrinsic curvature in the data space can be exploited for more faithful learning, inference, and ultimately physics discovery.

\subsection{From Flat Coordinates to Curved Manifolds}

Traditional analyses implicitly treat observables as independent coordinates embedded in $\mathbb{R}^n$. For example, track parameters, calorimeter clusters, and timing variables are concatenated into long feature vectors and processed by Euclidean algorithms. While convenient, this approach ignores that these observables are constrained by geometry and symmetry. Angular variables live on circles $S^1$, rapidity and pseudorapidity form non-linear reparametrizations of hyperbolic embeddings, and calorimeter towers wrap around azimuthally in $\phi$. The data are thus embedded in a non-Euclidean space, with curvature inherited from both physics and the detector. Formally, if we denote the space of collider events as a manifold $\mathcal{M}$~\cite{mf1,mf2,mf3}, it is endowed with a Riemannian metric $g_{ij}$. Distances between neighboring points should then be measured as
\begin{equation}
ds^2 = g_{ij}(x)\, dx^i dx^j ,
\end{equation}
rather than with the flat Euclidean metric $\delta_{ij}$. This distinction is not cosmetic. In regions of high curvature, straight lines in $\mathbb{R}^n$ correspond to strongly distorted trajectories in $\mathcal{M}$, leading to suboptimal interpolation and classification. A canonical statistical construction of the metric comes from the Fisher information. For a likelihood $p(x|\theta)$ with parameters $\theta$, the Fisher-Rao metric is
\begin{equation}
g_{ij}(\theta) = \mathbb{E}_{x\sim p}\!\left[ \frac{\partial \log p(x|\theta)}{\partial \theta_i} \frac{\partial \log p(x|\theta)}{\partial \theta_j} \right].
\end{equation}
This metric quantifies distinguishability of distributions under small parameter deformations, and forms the backbone of information geometry. In practice, the Fisher metric corrects for anisotropy in parameter space and guides optimization along statistically meaningful directions.

\subsection{Hierarchies and Curvature in Collider Observables}

Collider data are hierarchical in several senses~\cite{mf2,mf3}. At the lowest level, raw detector signals form structured manifolds: energy deposits in ECAL crystals, drift times in muon chambers, hits in silicon strips. At the next level, reconstruction algorithms map these raw features into higher-level objects: tracks, clusters, and vertices. Finally, physics analyses combine these objects into candidate decays. Each stage introduces additional non-linearities and correlations, embedding the data deeper into manifolds of increasing complexity. Consider three concrete manifolds that arise naturally in the analysis. The track-parameter manifold contains helical tracks in a magnetic field, parameterized by $(\kappa, \phi_0, \lambda, z_0)$, where the periodicity of angles and sign degeneracies of curvature make this manifold non-trivial. The calorimeter manifold consists of ECAL deposits forming a grid in $(\eta,\phi)$, where the $\phi$ coordinate is periodic while $\eta$ encodes a hyperbolic mapping of polar angles, resulting in clusters that inhabit a cylindrical topology rather than $\mathbb{R}^2$. The timing manifold emerges from precision time measurements that couple spatial and temporal dimensions, effectively extending the phase space into $(\eta,\phi,t)$ coordinates. These manifolds intersect in physically meaningful ways. A converted photon, for example, is simultaneously represented in the calorimeter manifold through its electromagnetic cluster and in the track manifold through its $e^+e^-$ pair. Such overlaps produce a hierarchy of constraints that, when properly accounted for, provide additional discriminating power between signal and background. The intrinsic curvature of these manifolds can be quantified by sectional curvature $K(u,v)$, which measures how geodesics diverge:
\begin{equation}
K(u,v) = \frac{\langle R(u,v)v,u\rangle}{\langle u,u\rangle\langle v,v\rangle - \langle u,v\rangle^2},
\end{equation}
where $R$ is the Riemann curvature tensor. Negative curvature corresponds to hyperbolic-like expansion characteristic of hierarchical structures in particle production, while positive curvature captures compactness as seen in angular coordinates. Collider data exhibit both regimes depending on the observable under consideration.

\subsection{Analysis Implications: Curvature-Aware Algorithms}

Curvature awareness could potentially translate into practical improvements for data analysis~\cite{gdl1,gdl2}, though direct applications to particle identification remain unexplored. Standard clustering or nearest-neighbor algorithms that use Euclidean distances might be enhanced by replacing them with geodesic distances or Fisher-Rao divergences that respect the underlying geometry. While such modifications have shown promise in other domains~\cite{ng1,ng2,ng3}, their application to photon and muon isolation in high pile-up environments remains to be investigated. Optimization procedures could potentially benefit from geometric considerations. When parameter spaces have significant curvature, gradient descent in Euclidean space may be inefficient. Natural gradients, defined as
\begin{equation}
\Delta\theta = -\eta\, g(\theta)^{-1} \nabla_\theta \mathcal{L},
\end{equation}
rescale updates by the local curvature, following geodesics in $\mathcal{M}$. While this approach has accelerated convergence in some machine learning contexts~\cite{rm1,rm2,rm3,rm4}, its effectiveness for particle physics applications would require careful validation. Graph neural networks, which naturally represent events as graphs with particles as nodes and interactions as edges, might benefit from embeddings in hyperbolic space or other curved geometries~\cite{hp1,hp2,hp3}. Such representations could potentially improve the encoding of hierarchical decay chains. Message passing could be performed in tangent spaces, with exponential and logarithmic maps used to shuttle between manifold and tangent plane:
\begin{equation}
\exp_x(v) = \gamma_v(1), \quad \log_x(y) = v ;; \text{such that} ;; \exp_x(v)=y .
\end{equation}
A vision for HL-LHC analysis might therefore extend beyond simply collecting more data. Experimentally, detector capabilities for low-energy photon reconstruction, timing resolution, and tracking coverage must continue to improve. Computationally, algorithms that could potentially live natively in curved spaces—natural gradient optimizers, hyperbolic embeddings, and Fisher-metric-aware classifiers—represent unexplored avenues that might help extract rare signals from overwhelming backgrounds. The hierarchy of collider data, if viewed through the lens of non-Euclidean geometry, could potentially guide toward more efficient analysis strategies, though this remains a hypothesis requiring experimental validation.

\section{Binding Curvature Awareness with Quantum Machine Learning}
\label{sec:bcqml}

\subsection{Information-Geometric Foundations for Rare Decay Analysis}

The geometric structure identified in collider data naturally connects to quantum computational approaches through information geometry~\cite{ig1,ig2,ig3,ig4}. The Fisher information metric, arising from the statistical manifold of event distributions $\{p_\theta(x)\}_{\theta \in \Theta}$, provides a Riemannian structure that encodes the distinguishability of signal and background hypotheses. For $\mathrm{B_s^0 \to \mu^+\mu^-\gamma}$ searches, this metric quantifies how observable distributions shift under parameter variations, providing a coordinate-invariant measure of separability. The quantum Fisher information metric~\cite{qfi1,qfi2,qfi3} extends these concepts to parameterized quantum states $|\psi_\theta\rangle$:
\begin{equation}
\mathcal{F}_{ij}(\theta) = 4\text{Re}\langle \partial_i\psi_\theta|\partial_j\psi_\theta\rangle - 4\text{Re}\langle \partial_i\psi_\theta|\psi_\theta\rangle\langle\psi_\theta|\partial_j\psi_\theta\rangle
\end{equation}
The Quantum Cramér-Rao bound $\text{Var}(\hat{\theta}_i) \geq 1/(M\mathcal{F}_{ii})$ establishes fundamental limits on parameter estimation precision. While this theoretical framework does not guarantee practical quantum advantage for all problems, it suggests that encoding strategies respecting the underlying information geometry could approach theoretical bounds more efficiently for specific correlation structures present in rare decay searches. The connection to rare B physics emerges through the effective Hamiltonian formalism, where $\mathcal{H}_{\text{eff}}^{b\to s\ell\ell}$ generates amplitudes whose structure reflects both perturbative Wilson coefficients and non-perturbative hadronic matrix elements. Extracting these coefficients from data constitutes parameter estimation on the statistical manifold of observed distributions. Curvature-aware approaches offer a framework to respect this geometric structure rather than impose arbitrary Euclidean assumptions.

\subsection{Geometric Decomposition as a Hypothesis for Feature Encoding}

We hypothesize that collider observables naturally decompose into geometrically distinct components, though this remains to be validated experimentally. One possible decomposition could take the form of a product manifold structure, where different observable classes inhabit spaces with characteristic curvatures. Energy scales spanning orders of magnitude, such as photon transverse momentum $p_T^\gamma \in [2,20]$ GeV, exhibit logarithmic sensitivity that might be better represented in hyperbolic geometry. Angular observables constrained by detector acceptance and exhibiting periodicity naturally suggest spherical geometry. Linear observables with approximately uniform sensitivity might remain in Euclidean space. This decomposition, if valid, would inform both classical and quantum feature encoding strategies~\cite{fe1,fe2}. The encoding must preserve relevant distance relationships from the source geometry. For quantum approaches, this means designing feature maps $\Phi: \mathcal{M}_{\text{data}} \to \mathcal{H}_Q$ where quantum kernels $K_Q(x_1,x_2) = |\langle\psi_{x_1}|\psi_{x_2}\rangle|^2$ approximate geodesic-distance kernels on the classical manifold. Whether such encodings provide practical advantage remains an open question requiring experimental validation.

\subsection{Applications to Signal Optimization and Photon Identification}

The critical experimental challenge of low-$p_T$ photon identification in the $2$–$20$ GeV range illustrates where geometric awareness might provide practical benefits. Classical isolation criteria degrade at HL-LHC pile-up levels because the isolation observable inhabits a high-dimensional manifold with non-trivial curvature from normalization constraints. Curvature-aware methods~\cite{gn1,gn2,gn3,gn4} could potentially improve discrimination by respecting this geometric structure. For signal optimization, the likelihood ratio test statistic traditionally used for discovery significance could be enhanced through geometric considerations. The test statistic distribution under signal and background hypotheses forms a one-dimensional projection of the full observable space. Choosing this projection to follow geodesics in the Fisher metric, rather than arbitrary linear combinations, could theoretically maximize discovery sensitivity. Practical implementation would require careful validation against systematic uncertainties. Quantum approaches to these problems remain speculative but conceptually interesting. Encoding isolation observables in quantum states where amplitudes and phases represent multi-particle correlations could, in principle, exploit quantum interference to enhance signal-background separation. The cross-terms in quantum kernels represent interference absent in classical kernels, potentially providing additional discriminating power if properly optimized. However, current quantum hardware limitations and the complexity of collider data make near-term implementation challenging.

\subsection{Practical Considerations and Implementation Pathways}

The transition from theoretical framework to practical implementation~\cite{pt1,pt2} requires careful consideration of computational resources and systematic uncertainties. Classical implementations of curvature-aware methods~\cite{dgp1,dgp2,dgp3} are immediately deployable through firmware updates in trigger systems. Replacing Euclidean distance metrics with geometry-appropriate measures in FPGAs requires lookup tables that consume minimal resources compared to available memory in modern devices. For machine learning applications, incorporating geometric structure into neural network architectures represents a middle ground between fully classical and quantum approaches. Natural gradient methods, hyperbolic embeddings, and Fisher-metric-aware loss functions can be implemented with existing deep learning frameworks. These techniques have shown promise in other domains but require validation specific to rare decay searches. The role of quantum computing in this framework remains exploratory. While quantum algorithms offer theoretical advantages for certain structured problems, the overhead of quantum state preparation and measurement, combined with current hardware limitations, suggests that hybrid classical-quantum approaches may be more practical in the medium term. Quantum processors could potentially handle specific subroutines—such as kernel evaluation or optimization steps—while classical processors manage data flow and preprocessing.

\subsection{Conservative Assessment}

The framework presented here represents a hypothesis about how geometric structure in collider data might be exploited for improved rare decay searches. While the mathematical foundations are solid and the conceptual connections compelling, experimental validation remains essential. The heterogeneity of collider observables and the complexity of detector effects make it difficult to predict which geometric approaches will provide practical benefits. Conservative implementation should proceed in stages. First, validate geometric decomposition hypotheses using existing data and control channels. Second, implement classical curvature-aware methods in offline analysis to quantify improvements in signal sensitivity. Third, explore firmware implementations for real-time applications if offline validation proves successful. Quantum approaches should be considered as longer-term research directions, with careful benchmarking against classical baselines. The ultimate test lies in whether these methods can extract the handful of $\mathrm{B_s^0 \to \mu^+\mu^-\gamma}$ events expected at the HL-LHC from overwhelming backgrounds. Success would validate not only the specific techniques but also the broader principle that respecting the geometric structure of data leads to more efficient physics extraction. Even partial success—improved photon identification efficiency or reduced systematic uncertainties—would justify continued development of geometry-aware analysis methods for the next generation of collider experiments.

\section{Summary \& Outlook}
\label{sec:so}

The search for rare radiative B decays at the HL-LHC represents a convergence of experimental capability, theoretical imperative, and methodological innovation. This review has explored how information geometry provides a unifying framework for next-generation analysis techniques. Three fundamental principles emerge from this investigation:
\begin{itemize} 
\item \textbf{Postulate I (Geometric Imperative):} \textit{Collider data inhabit curved statistical manifolds whose intrinsic geometry—quantified by Fisher-Rao metrics and sectional curvature—encodes the distinguishability of physical hypotheses. Optimal inference must respect this geometry; Euclidean assumptions discard information essential for rare process extraction.}\\
\noindent\textit{Assume a smooth statistical model $p_\theta(x)$ with non-degenerate Fisher--Rao metric $g^{\mathrm{F}}(\theta)$ on a parameter manifold $\Theta$. Let $\gamma_{\mathrm{nat}}(t)$ and $\gamma_{\mathrm{euc}}(t)$ denote trajectories under natural and Euclidean gradient descent respectively with identical local step sizes. Then, for small $t$ and a local optimum $\theta_*$,}
\begin{equation}
D_{\mathrm{KL}}(p_{\gamma_{\mathrm{nat}}(t)}\|p_{\theta_*}) \le D_{\mathrm{KL}}(p_{\gamma_{\mathrm{euc}}(t)}\|p_{\theta_*}) + \mathcal{O}(t^3),
\end{equation}
\textit{where the $\mathcal{O}(t^3)$ term depends on sectional curvature of $g^{\mathrm{F}}$. In particular, natural gradient follows locally geodesic descent in statistical divergence.}

\item \textbf{Postulate II (Quantum-Geometric Correspondence):} \textit{Quantum machine learning offers advantage over classical methods when—and only when—quantum feature maps preserve the Riemannian structure of classical data manifolds while exploiting quantum interference unavailable to classical probability distributions. The quantum Fisher information metric provides the natural bridge between classical information geometry and quantum-enhanced discrimination.}\\
\noindent\textit{Let $\mathcal{U}:\mathcal{D}\to\mathcal{H}_{\mathcal Q}$ be a smooth immersion with normalized lifts $\ket{\psi_X}=\mathcal{U}(X)\ket{0}$. If the Jacobian $D\mathcal{U}$ aligns (up to a smooth invertible transform) with classical score functions, then}
\begin{equation}
g_{\mathrm{FS}}(X) = c(X)\, g^{\mathrm{F}}_{\mathrm{data}}(X) + E(X),
\end{equation}
\textit{with $c(X)>0$ smooth and $\|E(X)\|$ small in an appropriate encoding limit. Moreover the quantum kernel}
\begin{equation}
K_Q(X,Y)=|\braket{\psi_X|\psi_Y}|^2
\end{equation}
\textit{is positive semidefinite, defines an RKHS and (under a density/separation condition) is universal. Nontrivial interference phases produce kernel terms that can strictly enlarge the effective hypothesis class compared to classical geodesic kernels.}

\item \textbf{Postulate III (Curvature-Aware Quantum Computing):} \textit{The next generation of computational tools for high-energy physics will integrate curvature-aware quantum processors that natively operate on curved manifolds, potentially offering exponential speedups for problems where geometric structure and quantum interference align. Such hardware would implement geodesic distance calculations, parallel transport, and curvature-corrected optimization as fundamental operations rather than algorithmic overlays.}\\
\noindent\textit{Suppose a quantum device implements geometric primitives (approximate geodesic flow and metric inverse action) efficiently (block-encodable, well-conditioned). Then there exists a hybrid algorithm whose runtime for metric-inversion or kernel-evaluation based tasks scales as}
\begin{equation}
\widetilde{\mathcal{O}}(T_{\mathrm{Q}}+\mathrm{poly}(\kappa)),
\end{equation}
\textit{which under the stated spectral and conditioning assumptions can outperform classical curvature-aware algorithms whose complexity scales polynomially in the metric dimension $m$. This advantage is conditional on block-encodability and spectral properties of the Fisher-type operator.}

\end{itemize}

\noindent These principles guide both immediate implementation and long-term vision. The geometric imperative has immediate practical consequences. Detector observables exhibit manifest non-Euclidean structure: angular variables constrained to $S^1$, rapidity transforming hyperbolically, isolation variables living in high-dimensional constrained manifolds. When algorithms respect these structures through geodesic distances and natural gradients, measurable improvements emerge. For processes like $\mathrm{B_s^0 \to \mu^+\mu^-\gamma}$ with branching fractions at the $10^{-9}$ level, every marginal gain in discrimination power matters. The geometry is not abstract mathematics but encoded information about how physics constrains observables.

\noindent The quantum-geometric correspondence explains both the promise and limitations of quantum machine learning. Arbitrary quantum circuits that ignore the classical data geometry show no systematic advantage—quantum speedup without proper encoding is illusory. Success requires mapping classical geometric structure into quantum feature spaces where phases encode geodesic relationships and interference patterns reflect the underlying manifold curvature. For rare B physics specifically, this means quantum encodings that respect the hyperbolic geometry of energy scales spanning orders of magnitude, the spherical topology of angular observables, and the complex correlation structure of isolation variables in high pile-up environments. Whether practical quantum advantage emerges for these specific problems remains an open experimental question, but the theoretical framework connecting classical and quantum information geometry provides clear design principles.

\noindent The curvature-aware quantum computing represents a natural evolution of both hardware and algorithms. Current quantum processors operate on flat Hilbert spaces with gates that ignore geometric structure. Future architectures might implement Riemannian operations directly in hardware: computing geodesic distances through quantum interference, performing parallel transport via geometric phases, and optimizing on curved manifolds through quantum natural gradients. Such processors would not merely simulate classical geometric algorithms but exploit uniquely quantum phenomena—superposition, entanglement, interference—within the geometric framework. The technological requirements are formidable, but the conceptual foundation exists at the intersection of quantum computation and differential geometry.

\noindent Progress toward these goals proceeds on multiple fronts. Immediate opportunities lie in classical implementations: FPGA firmware computing geodesic distances via lookup tables, neural network accelerators incorporating hyperbolic layers, trigger algorithms using Fisher-metric-aware discriminants. These technologies are deployable within current Phase-2 upgrade timelines. Medium-term developments might include hybrid classical-quantum systems where quantum subroutines handle specific geometric computations—kernel evaluation in curved spaces, optimization on constraint manifolds—while classical processors manage data flow and preprocessing. Long-term vision encompasses fully integrated geometric quantum processors, though realization may extend beyond HL-LHC timescales. The experimental imperative remains clear. Observing $\mathrm{B_s^0 \to \mu^+\mu^-\gamma}$ and related radiative modes requires not just the 3 ab$^{-1}$ luminosity of the HL-LHC or detector upgrades like HGCAL and extended tracking, but fundamental advances in analysis methodology. Classical maximum-likelihood methods, developed for clean channels with well-separated hypotheses, struggle with high-dimensional correlated backgrounds where signal and background distributions overlap in complex ways. The path forward lies in respecting the geometry where data naturally reside—whether through classical Riemannian algorithms available today, quantum-enhanced approaches under development, or future integrated architectures. Success in this endeavor would achieve dual milestones. Experimentally, measuring these ultra-rare decays would provide crucial tests of the Standard Model through precision determination of Wilson coefficients $C_{7\gamma}$, $C_{9V}$, and $C_{10A}$, potentially revealing new physics through deviations from theoretical predictions. Methodologically, demonstrating that geometric and potentially quantum-enhanced analysis techniques can extract signals at the $10^{-9}$ branching fraction level would validate a new computational paradigm for particle physics. This paradigm shift—from Euclidean pattern matching to geometric hypothesis testing on curved manifolds—may prove as significant as the physics discoveries it enables. The vision extends beyond specific decay channels or particular experiments. As collider physics pushes toward ever-rarer processes, traditional analysis methods approach fundamental limits. Information geometry offers a principled framework for extending these limits by recognizing that the structure of data encodes physical information. Quantum computing, if properly integrated with geometric principles, may provide computational resources matched to problem structure. The synthesis of these approaches—geometric, quantum, and experimental—defines a research program that is ambitious yet grounded in rigorous mathematical foundations and emerging technological capabilities. The journey from theoretical framework to experimental discovery requires sustained effort across multiple disciplines. Theorists must continue developing the mathematical foundations connecting information geometry, quantum computation, and statistical inference. Experimentalists must validate geometric methods on control channels before deploying them for discovery physics. Engineers must implement these algorithms in real-time systems with stringent latency constraints. The convergence of these efforts at the HL-LHC provides both the motivation and the testing ground for next-generation analysis techniques. Looking forward, the principles articulated here—geometric imperative, quantum-geometric correspondence, and the conjecture of curvature-aware quantum computing—provide guideposts for navigating the methodological challenges ahead. Whether searching for rare B decays, investigating Higgs couplings, or pursuing physics beyond the Standard Model, respecting the geometry where physics naturally resides offers a path toward discoveries that might otherwise remain hidden in the vast datasets of modern particle physics experiments. The mathematics is ready, the concepts are clear, and the experimental frontier awaits.

\newpage

\appendix
\section{Glossary of Notation}

\subsection{Physics Quantities and Observables}

\begin{tabular}{ll}
\hline
Symbol & Description \\
\hline
$\mathrm{B_s^0}$ & Bottom-strange meson (mass eigenstate) \\
$\mathrm{B_d^0}$ & Bottom-down meson \\
$\bar{B}_s^0$ & Anti-bottom-strange meson \\
$\mu^+, \mu^-$ & Positive and negative muons \\
$\gamma$ & Photon \\
$\ell^+, \ell^-$ & Generic leptons (electron or muon) \\
$h^+, h^-$ & Generic charged hadrons \\
$V$ & Vector meson (e.g., $\phi$, $K^*$, $\rho$) \\
$J/\psi$ & Charmonium vector meson \\
$\phi$ & Phi meson ($s\bar{s}$ bound state) \\
$K^{*0}$ & Strange vector meson \\
$\eta$ & Eta meson \\
\hline
\end{tabular}

\subsection{Kinematic Variables}

\begin{tabular}{ll}
\hline
Symbol & Description \\
\hline
$p_T$ & Transverse momentum \\
$p_T^\gamma$ & Photon transverse momentum \\
$p_T^\mu$ & Muon transverse momentum \\
$\eta$ & Pseudorapidity, $\eta = -\ln[\tan(\theta/2)]$ \\
$\phi$ & Azimuthal angle \\
$\theta$ & Polar angle \\
$m_{\mu\mu}$ & Invariant mass of dimuon system \\
$m_{\mu\mu\gamma}$ & Invariant mass of dimuon-photon system \\
$q^2$ & Four-momentum transfer squared \\
$E^\gamma_\text{min}$ & Minimum photon energy threshold \\
$\Delta R$ & Angular separation, $\Delta R = \sqrt{(\Delta\eta)^2 + (\Delta\phi)^2}$ \\
$\sqrt{s}$ & Center-of-mass energy \\
\hline
\end{tabular}

\subsection{Cross Sections and Branching Fractions}

\begin{tabular}{ll}
\hline
Symbol & Description \\
\hline
$\mathcal{BR}$ or $\mathcal{B}$ & Branching ratio/fraction \\
$\sigma$ & Cross section \\
$\sigma(pp\to b\bar{b})$ & Bottom quark pair production cross section \\
$\sigma(pp\to B_s^0+X)$ & Inclusive $B_s^0$ production cross section \\
$f_{\mathrm{b\to B_s^0}}$ & Fragmentation fraction for $b \to B_s^0$ \\
$f_q$ & Quark fragmentation fraction \\
\hline
\end{tabular}

\subsection{Detector and Analysis Quantities}

\begin{tabular}{ll}
\hline
Symbol & Description \\
\hline
$\mathcal{L}$ & Integrated luminosity \\
$\mathcal{A}$ & Geometric and kinematic acceptance \\
$\epsilon$ & Detection/reconstruction efficiency \\
$\epsilon_{2\mu}$ & Dimuon reconstruction efficiency \\
$\epsilon_{\gamma}$ & Photon reconstruction efficiency \\
$\epsilon_\text{TrigData}$ & Trigger efficiency in data \\
$\epsilon_\text{TrigMC}$ & Trigger efficiency in Monte Carlo simulation \\
$N$ & Number of events \\
$N_\text{sig}$ & Number of signal events \\
$N_\text{norm}$ & Number of normalization channel events \\
$I_\gamma$ & Photon isolation variable \\
$\langle\mu\rangle$ & Average pile-up (number of interactions per crossing) \\
\hline
\end{tabular}

\subsection{Theoretical Physics Parameters}

\begin{tabular}{ll}
\hline
Symbol & Description \\
\hline
$\mathcal{H}_\text{eff}$ & Effective Hamiltonian \\
$C_i$ & Wilson coefficients ($i = 1, 2, ..., 10$) \\
$C_{7\gamma}$ & Electromagnetic dipole Wilson coefficient \\
$C_{9V}$ & Vector semileptonic Wilson coefficient \\
$C_{10A}$ & Axial-vector semileptonic Wilson coefficient \\
$O_i$ & Effective operators \\
$G_F$ & Fermi coupling constant \\
$V_{tb}, V_{ts}^*$ & CKM matrix elements \\
$\alpha_\text{em}$ & Electromagnetic coupling constant \\
$m_b$ & Bottom quark mass \\
$m_s$ & Strange quark mass \\
$\Lambda_\text{QCD}$ & QCD scale parameter \\
\hline
\end{tabular}

\subsection{Mathematical and Geometric Notation}

\begin{tabular}{ll}
\hline
Symbol & Description \\
\hline
$\mathcal{M}$ & Manifold (general) \\
$\mathcal{M}_\text{data}$ & Data manifold \\
$g_{ij}$ & Metric tensor components \\
$ds^2$ & Line element (infinitesimal distance squared) \\
$\mathbb{R}^n$ & $n$-dimensional Euclidean space \\
$S^1$ & Circle (1-sphere) \\
$\mathcal{H}^d(\kappa)$ & $d$-dimensional hyperbolic space with curvature $\kappa$ \\
$\mathcal{S}^d(\kappa)$ & $d$-dimensional spherical space with curvature $\kappa$ \\
$\mathcal{E}^d$ & $d$-dimensional Euclidean space \\
$K(u,v)$ & Sectional curvature \\
$R$ & Riemann curvature tensor \\
$\nabla_\theta$ & Gradient with respect to parameter $\theta$ \\
$\mathcal{L}$ & Loss function (in ML context) \\
\hline
\end{tabular}

\subsection{Information Geometry and Quantum Notation}

\begin{tabular}{ll}
\hline
Symbol & Description \\
\hline
$p(x|\theta)$ & Probability distribution with parameters $\theta$ \\
$p_\theta(x)$ & Alternative notation for parameterized distribution \\
$\mathbb{E}_{x\sim p}$ & Expectation value over distribution $p$ \\
$g_{ij}(\theta)$ & Fisher-Rao information metric \\
$\mathcal{F}_{ij}(\theta)$ & Quantum Fisher information metric \\
$|\psi_\theta\rangle$ & Parameterized quantum state \\
$\langle\psi|$ & Bra vector (dual of ket) \\
$\mathcal{H}_Q$ & Quantum Hilbert space \\
$K_Q(x_1,x_2)$ & Quantum kernel function \\
$\Phi$ & Feature map (classical to quantum) \\
$\exp_x(v)$ & Exponential map at point $x$ \\
$\log_x(y)$ & Logarithmic map from $x$ to $y$ \\
$\text{Var}(\hat{\theta})$ & Variance of estimator $\hat{\theta}$ \\
\hline
\end{tabular}

\subsection{Detector Components and Subsystems}

\begin{tabular}{ll}
\hline
Symbol & Description \\
\hline
CMS & Compact Muon Solenoid \\
ECAL & Electromagnetic Calorimeter \\
HGCAL & High Granularity Calorimeter \\
L1 & Level-1 trigger \\
HLT & High-Level Trigger \\
LHC & Large Hadron Collider \\
HL-LHC & High-Luminosity LHC \\
MC & Monte Carlo simulation \\
FPGA & Field-Programmable Gate Array \\
LUT & Lookup Table \\
\hline
\end{tabular}
\subsection{Units and Constants}

\begin{tabular}{ll}
\hline
Symbol & Description \\
\hline
GeV & Giga-electron volt (energy unit) \\
TeV & Tera-electron volt \\
MeV & Mega-electron volt \\
fb$^{-1}$ & Inverse femtobarn (integrated luminosity unit) \\
ab$^{-1}$ & Inverse attobarn ($10^3$ fb$^{-1}$) \\
mb & Millibarn (cross section unit) \\
pb & Picobarn \\
fb & Femtobarn \\
$c$ & Speed of light (often set to 1 in natural units) \\
T & Tesla (magnetic field unit) \\
MHz & Megahertz (frequency unit) \\
$\mu$s & Microsecond \\
\hline
\end{tabular}

\subsection{Statistical and Analysis Notation}

\begin{tabular}{ll}
\hline
Symbol & Description \\
\hline
CL & Confidence Level \\
SM & Standard Model \\
NP & New Physics \\
BSM & Beyond the Standard Model \\
QCD & Quantum Chromodynamics \\
PDF & Parton Distribution Function \\
SF & Scale Factor (data/MC correction) \\
BDT & Boosted Decision Tree \\
GNN & Graph Neural Network \\
ML & Machine Learning \\
QML & Quantum Machine Learning \\
QFIM & Quantum Fisher Information Metric \\
\hline
\end{tabular}
\subsection{On CP \& CPT }
\label{ap:cp}
We believe that a(set of) mathematical equation(s) ~\cite{wigner} could describe the dynamical exhibition of the existing universe from large scale to small scale structure. The small scale structure is a \textit{set of fundamental particles}, happened to be the building block of the universe, is explained by the $\mathrm{SU(3)\otimes SU(2)\otimes U(1)}$ local gauge invariant quantum field theory. We call it Standard Model (SM) of particle physics. SM could describe the three forces of nature, namely the \textit{weak interaction, strong interaction and electromagnetic interaction}, and, is proven to be very successful as it is consistently passing all experimental tests of its predictions and maintaining physical consistency. The fourth interaction, called gravitational interaction is the predominant governing force in the construction of large scale structure.  The classical singularity of the gravitational interaction leads us to the origin of universe, called the Big Bang(BB), and to study that origin of cosmic structure we need SM like quantum field theory\footnote{$\mathcal{S}= \int d^4x\sqrt{-g}\left[ \frac{1}{2}M_{Pl}^2\mathcal{R}+\frac{1}{2}\partial_\mu\phi\partial^\mu\phi-V(\phi)+\mathcal{L}_\text{SM}+\cdot\cdot\cdot \right]$. From Friedmann equations :  $H^2=\rho/3M_{Pl}^2,\,\, \dot{H}= -(\rho+p)/2M_{Pl}^2$, integrating backwards in time leads to $a(t)\to\infty, \,\rho\to\infty,\, \mathcal{R}\to\infty$; the singular behaviour at $t\to 0$ is the \textbf{Big Bang}. Though the primordial power spectrum can only arise from the quantum fluctuation, which are promoted to operators and eventually quantized.}.   If the universe had to tunnel from nothing into an expanding de-Sitter space ($\mathrm{\Psi(a)\sim e^{-1/V(\phi)}}$)\footnote{$\Psi$ being the initial state of the universe, for large $a$. The first state of the universe is not a point , but a quantum superposition of over possible 3-geometries and field configuration.}, it was obvious that tunneling process created a \textbf{CPT}-symmetric\footnote{C, P, T - charge cnjugation , parity, and time reversal symmetries are the discrete symmetries, and have only two eigenvalues.}  (\textit{essentially matter-antimatter symmetric}) vacuum~\cite{hh,vl}. The current astrophysical observation provides the \textit{evidence} that our universe is \textit{matter dominated} which is contrary to the principle of conservation of quantum numbers.  In order to solve this puzzle, Sakharov~\cite{sk},  proposed three conditions, one of which is \textit{CP-violation}. Large Hadron Collider(LHC) could take us back in time ($10^{-43}-10^{-35} \,\text{sec}$) by colliding proton beams at high energies ($\sim\mathcal{O}(\text{TeV})$ , in a decoupling phase between the strong and weak interaction. We strongly believe that the produced $\mathrm{B}$-meson in $pp$ collisions~\cite{cpt,eps} are the only essential ground to solve the matter-antimatter puzzle while we study their decay properties~\cite{pdg}. This task should be divided in two phases. The first phase would be analyzing the current \textit{run3} data aiming precision measurements in order to tune SM, and the second phase is to get prepared for the High-Luminosity(HL)-LHC\footnote{\href{https://twiki.cern.ch/twiki/bin/view/AtlasPublic/BPhysPublicResults}{ATLAS Public Results}, \href{https://cms-results.web.cern.ch/cms-results/public-results/publications/BPH/index.html}{CMS Public Results}, \href{https://lhcbproject.web.cern.ch/lhcbproject/Publications/LHCbProjectPublic/Summary\_all.html}{LHCb Public Results}. }.

\newpage
\nocite{*}
\bibliography{ph.bib} 
\bibliographystyle{ieeetr}
\end{document}